\documentclass{aa}  

\usepackage{graphicx}
\usepackage{txfonts}
\usepackage{amstext}

\begin{document} 

\titlerunning{Analysis of putative exoplanetary signatures found in two sdBVs}
   \title{Analysis of putative exoplanetary signatures found in light curves of two sdBV stars observed by \textit{Kepler}}
   \subtitle{}

 \author{A.~Blokesz\inst{1}\and J.~Krzesinski\inst{1,2} \and L.~Kedziora-Chudczer\inst{3}}

   \institute{Mt. Suhora Observatory, Pedagogical University of Cracow,
        ul. Podchor\c{a}\.{z}ych 2, 30-084 Cracow, Poland
              \\
              \email{abl@astro.as.up.krakow.pl}, \email{jk@astro.as.up.krakow.pl}, \email{lkedzior-at-unsw.edu.au}
             \and
             Astronomical Observatory, Jagiellonian University, ul. Orla 171, PL-30-244 Krakow,
             Poland
             \and
             University of New South Wales, School of Physics and Astrophysics,
             Sydney NSW 2052, New South Wales, Australia 
             }
   \date{Received 2018}

 
  \abstract
   {We investigate the validity of the claim that invokes two extreme exoplanetary system candidates around the pulsating B-type subdwarfs KIC\,10001893 and KIC\,5807616 from the primary $\it{Kepler}$ field.
   }
   {Our goal was to find characteristics and the source of weak signals that are observed in these subdwarf light curves.
   }
   {To achieve this, we analyzed short- and long-cadence $\it{Kepler}$ data of the two stars by means of a Fourier transform and compared the results to Fourier transforms of simulated light curves to which we added exoplanetary signals. The long-cadence data of KIC\,10001893 were extracted from CCD images of a
nearby star, KIC\,10001898, using a point spread function reduction technique.}
   {It appears that the amplitudes of the Fourier transform signals that were found in the low-frequency region depend on the methods that are used to extract and prepare $\it{Kepler}$ data. We demonstrate that using a comparison star for space telescope data can significantly reduce artifacts. Our simulations also show that a weak signal of constant amplitude and frequency, added to a stellar light curve, conserves its frequency in Fourier transform amplitude spectra to within 0.03\,$\mu$Hz.}
   {Based on our simulations, we conclude that the two low-frequency Fourier transform signals found in KIC\,5807616 are likely the combined frequencies of the lower amplitude pulsating modes of the star. In the case of KIC\,10001893, the signal amplitudes that are visible in the light curve depend on the data set and reduction methods. The strongest signal decreases significantly in amplitude when KIC\,10001898 is used as a comparison star. Finally, we recommend that the signal detection threshold is increased to 5\,$\sigma$ (or higher) for a Fourier transform analysis of $\it{Kepler}$ data in low-frequency regions.}


   \maketitle

%

\section{Introduction}

It has recently been claimed that the two pulsating subdwarf B (sdBV) stars KIC\,5807616 ($T_{eff}$=\,27\,730\,K, $\log$\,g\,=\,5.52) \citep{Charpinet2011} and KIC\,10001893 ($T_{eff}$ = 27\,500 K, $\log$\,g\,=\,5.35) \citep{Silvotti2014} in the primary $\it{Kepler}$ \citep{KepTel2016} field harbor planetary candidate companions. The claims were based on signals that were found in the low-frequency region of the Fourier transform (FT) of the sdBV light curves. It was shown that no pulsation frequency combinations of high-amplitude gravity (g) or pressure (p) modes that were identified for these stars could cause FT low-frequency signals. Because g mode pulsations cannot occur below $\sim$60\,$\mu$Hz in these stars, that is, beyond the cutoff frequency  \citep{Hansen1985} because of severe pulsation damping, the authors inferred an exoplanetary origin of the signals. 
In the case of KIC\,10001893, the situation was more complicated because pulsating modes of the star do not show multiplet structures in the FT of the stellar light curve. This means that the star is either a slow rotator or that its pulsation axis is inclined toward the observer \citep{Charpinet2011}. 
Based on arguments of orbital stability, \cite{Silvotti2014} constrained the 
inclination of the system to be greater than 1 to 3 degrees, depending on the 
mean density of the planets.

The interpretations and reasonings by \cite{Charpinet2011} and \cite{Silvotti2014} are very appealing, but there are some arguments against them. The sdBVs are extreme horizontal branch stars in the Hertzsprung-Russel (HR) diagram. These are hot and low-mass stars ($\sim$0.5 solar masses, $\sim$0.2 solar radii, and a $T_{eff}$ between 20\,000 - 40\,000~K) that burn helium in their cores. They lost most of their envelope hydrogen in  the red giant (RG) stage before the helium flash, which means that their evolutionary tracks entirely omit the asymptotic giant branch and that the star directly moves to the white dwarf cooling track \citep{Heber2016}. 
Exoplanets orbiting an sdBV would have had to survive the RG phase of the parent star. An additional argument against planetary survivors of the RG phase is the lack of exoplanet detections around white dwarfs, which are the successors of subdwarfs. However, it is not certain that planets around single white dwarfs are absent or rare. The existence of metal-polluted white dwarfs may be an indirect indication of exoplanets. 
Recent observations of WD 1145+017 (one of many metal-polluted white dwarfs) by \citet[and references therein]{Vander2018} suggests that there may be disintegrating planetary debris from a former exoplanetary system.
It is argued that if planets were to survive the giant phase of the parent
star, they would later perturb orbits of asteroids so that rocky material is 
accreted onto the white dwarf \cite[e.g., ][]{Jura2014}. This would be visible as 
metal lines in white dwarf spectra.

With the application of different data-reduction techniques and light-curve simulations, we are able to describe low-frequency signals in a qualitative way. This involves methods for estimating the amplitude error of FT signals and the stability of a signal frequency over time in the running FT. Our approach to the low-frequency signal analysis allows us to distinguish between constant exoplanetary and non-exoplanetary signals, as we show in this work.

\section{\textit{Kepler} data}
 
The data we used for our analysis were collected from the Barbara A. Mikulski Archive for Space Telescopes. For the sdBV KIC\,5807616, we used light curves that were extracted from short-cadence (SC) Q\,5\,-\,Q\,17 pixel data ($\it{Kepler}$ data are divided into 90-day (Q) quarters, \cite{Murphy2012}), the same as described in \cite{Krzesinski2015}. 
In the case of sdBV KIC\,10001893, the archive contains two sets of $\it{Kepler}$ data from which the stellar light curves can be extracted. The light curve that was analyzed by \cite{Silvotti2014} consists of Q\,3.2, Q\,6\,-\,Q\,17.2 SC, Q\,4, and Q\,5 long-cadence (LC) pixel data of KIC\,10001893 (referred to below as \textit{\textup{mixed data}}). The second light curve that we mainly use here is composed of continuous LC Q\,1\,-\,Q\,17.2 pixel data of KIC\,10001898, which is a bright nearby star to the KIC\,10001893 variable.

The significance of bright companion data is that they contain entire images of the KIC\,10001893 sdBV throughout the entire Q1\,-\,Q\,17.2 observing time. Consequently, the light curves of both stars could be derived from KIC 10001898 LC pixel data. To achieve this, we used two CCD data reduction techniques that are equivalent to point spread function (PSF) profile fitting and aperture photometry (see kepprf, kepcotrend, and kepextract task documentation of the PyKE command-line tools\footnote{  https://keplerscience.arc.nasa.gov/software.html}). Because of the \textit{Kepler} CCD pixel size of 3.98 arcsec and the 8.5 arcsec separation of both KIC\,10001893 sdBV and its bright companion ($\it{Kepler}$ magnitudes of 15.85 and 14.98, respectively), light contamination was a serious problem. The apertures were therefore composed of manually selected pixels, and we avoided pixels that were adjacent to a bright neighbor. Next, the aperture-reduced data were preprocessed using co-trending basis vectors (see $\it{Kepler}$ Data Release 12 Notes). The PSF and aperture flux data were d-trended using three-day spline functions. This resulted in the strong reduction of amplitudes for light-curve FT frequency peaks below $\sim$1-2\,$\mu$Hz. All light curves were examined, and outliers above 3\,$\sigma$ from the ten-point running mean were removed. Finally, the fluxes were converted into parts per million (ppm) units. 

To analyze the frequency combination (in the next section), we also used aperture-reduced SC and mixed~data
that were processed in similar ways. We did not use the standard (aperture) $\it{Kepler}$ pipeline fluxes for KIC\,10001893 because light contamination from the nearby star is strong. 
For the target, the contamination was typically responsible for quenching the observed pulsation amplitudes by about 17$\%$ (compared to the amplitudes derived from PSF fluxes). The highest contamination occurs in quarters 4,\,8,\,12, and 16, where both star images appear to be in contact as a result of an unfortunate $\it{Kepler}$ CCD orientation. During these quarters, the pulsation amplitudes are reduced by up to 50$\%$.

\begin{figure}
        \centering
        \includegraphics[width=8.8cm, clip=]{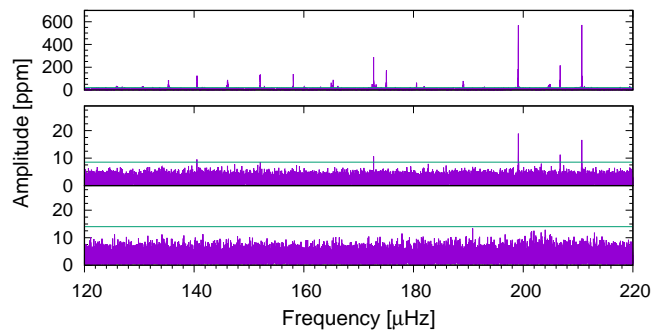}
        \caption{Fourier transforms of the bright neighbor KIC\,10001898 and sdBV KIC\,10001893 light curves extracted from LC data. The pulsation pattern of KIC\,10001893 sdBV (top panel, PSF data FT) is clearly visible in the aperture data for the bright neighbor (middle panel). The FT of the PSF\ data for the bright neighbor (bottom panel) shows no sign of pulsations from the sdBV. Green lines show the 4$\sigma$ signal detection thresholds of the light curve FTs. The amplitude scale changes in the two bottom panels.}
        \label{Fig_1}
\end{figure}

The contamination problem is shown in Fig.\,\ref{Fig_1}, where we show flux FTs for LC Q1-Q17.2 data that were reduced with both techniques. The top panel presents the FT of the PSF light curve of KIC\,10001893, while the bottom panels show
FTs of the bright companion light curves reduced with aperture and PSF methods. 
The stellar pulsation pattern is clearly visible in the top panel. 
Green horizontal lines indicate the 4\,$\sigma$ detection thresholds of the light curve FTs that were calculated within the 120\,-\,220\,$\mu$Hz range. They are close to 27\,ppm for the variable light-curve FT (barely visible because of the scale of the top panel), and 9.0 and 13.7\,ppm for the aperture and PSF data, respectively, for the bright neighbor star.
The FT of the aperture data is contaminated by the signal of the fainter sdBV star. Only the PSF data of the bright neighbor are free of the pulsation pattern of KIC\,10001893.   
 
At the basic level, SC and LC data are built from the same 6.02\,s exposures, 
followed by the 0.52\,s readout and added up, to produce 58.84876\,s SC 
data, versus the 29.4244 minute LC integrations \citep{Gilliland2010}. 
The difference is in the limiting Nyquist frequency (equal to half the rate 
at which a signal is being sampled), which in the case of LC data limits the FT 
frequency analysis to 283.21\,$\mu$Hz. At this limit, pulsation amplitudes in 
FT spectrograms are strongly damped compared to the FT of SC amplitudes 
(35\,\% at $\sim$270\,$\mu$Hz, and 20\,\% at $\sim$170\,$\mu$Hz). Near the low-frequency FT 
region, the differences are negligible. Below $\sim$20\,$\mu$Hz, LC 
data have more high-amplitude artifacts than SC data. However, in the 
frequency region between $\sim$20 and 60\,$\mu$Hz, the two spectra of the FT amplitude converge and are almost the same, regardless of the choice of the 
data cadence. Therefore, the longer homogeneous LC data data might be a better 
choice for a low-frequency analysis than the SC or mixed data. 

\section{Frequency combinations}   

\begin{table*}[t]
        \caption{KIC\,5807616: Frequency combinations that fall in the proximity of the $F_{1}$ and $F_{2}$ signals (here 48.182\,$\mu$Hz and 33.839\,$\mu$Hz, respectively).
        The frequencies with the highest amplitudes, best-fitting to the $F_{1}$ and $F_{2}$ signals
         are printed in bold; $f_{i}$ and $f_{j}$ are the prewhitened pulsation frequencies.}             
        \label{combi_char}      
        \centering                          
        \begin{tabular}{l r r c c}        
                \hline\hline                 
                Data & ${~f_{i}}~-~{~f_{j}}$~~~~~~~~~~ & $f_{i}, f_{j}$ ampl. & combination & combination to $F_{l}$\\  
                & [$\mu$$Hz$]~~~~~~~~~~~~ & [$ppm$]~~~~   &    [$\mu$$Hz$]     &
                distance [$\mu$$Hz$] \\
                \hline
                
                
                \\
                \smallskip
                SC & $ \bf{201.667 ~ - 167.848}~\,$ & 125~~,~~~836 & 33.818 &  $ \bf{0.021}$ $(F_2)$ \\
                & ${201.751}~ - 167.901~\,$ &  142~~,~~~353 & 33.851 & 0.012  $(F_2)$\\
                \\
                & $\bf{167.693}~ - 119.526~\,$ &  239~~,~~~~49 & 48.168 & $\bf{0.014}$  $(F_1)$\\
                & ${249.824}~ - 201.667~\,$ &  95~~,~~~125 &  48.157 & 0.024 $(F_1)$\\
        \end{tabular}
\end{table*}

Many pulsating stars at different evolutionary stages frequently exhibit combination modes in FTs of their light curves. For two pulsating modes $f_{i}$ and $f_{j}$ of high enough amplitudes, the nonlinear pulse-shape distortion will result in harmonics and combination frequencies being present in their FT spectrum. To the second order of perturbation, FT peaks at beat frequency $|f_{i} - f_{j}|$, at second harmonics $2f_{i}$, $2f_{j}$, and at sums of $f_{i}$ and $f_{j}$ can therefore be observed. The higher order perturbations will cause more combinations to appear. 

\begin{table*}[t]
        \caption{KIC\,10001893: Frequency combinations of pulsating modes (Col.\,2) derived from various types of data (Col.\,1), where data types are: LC (PSF long-cadence light curve ratio of $sdBV$ and the bright neighbor; see Sect. 4 for explanations), SC (aperture $sdBV$ light curve) and UZ \citep[frequencies from ][]{Uzundag2017}. Col.\,2 in round brackets: IDs of the closest match from the \cite{Uzundag2017} frequency list. Col.\,3: frequency amplitudes from Col.\,2. Col.\,4: Combination frequency values. Col.\,5: Distances of $k_{1}$,\,$k_{2}$, and\,$k_{3}$ from the frequencies shown in Col.\,4. Only pulsating modes with amplitudes greater than 30\,ppm are shown.}             
        \label{combi_sil}      
        \centering                          
        \begin{tabular}{l r r c c}        
                \hline\hline                 
                Data & $n~\cdot{~f_{i}}~-~m~\cdot{~f_{j}}$~~~~~~~~~~~~~~~~ & $f_{i}, f_{j}$ ampl. & combination & combination to $k_{l}$\\  
                & [$\mu$$Hz$]~~~~~~~~~~~~~~~~~~~~~~~~~ & [$ppm$]~~~~   &    [$\mu$$Hz$]     &    
                distance [$\mu$$Hz$] \\
                \hline
                
                \\
                \smallskip
                LC & $ \bf{238.9865 ~(f_{\,62}) - \,~~~~224.7199 ~(f_{\,59})}~\,$ & 71~~,~~~38 & 14.2666 &  $ \bf{0.0056 (k_3)}$ \\ 
                & $3\cdot{158.0832}~(f_{35}) - 3\cdot{140.5211}~(f_{29})~\,$ & 155~~,~138 & 52.6862 &  0.0039 $(k_1)$ \\
                
                \\
                \smallskip
                SC & $ \bf{238.9855 ~(f_{62}) -  \,~~~~224.7196 ~(f_{\,59})}~\,$ & 73~~,~~~54 & 14.2659 &  $ \bf{0.0049~(k_3)}$ \\
                & $2\cdot{114.6710}~(f_{18}) - \,~~~~176.6694~(f_{44})~\,$ & 34~~,~~~34 & 52.6727 &  0.0096 $(k_1)$ \\ 
                & $2\cdot{132.4709}~(f_{27}) - 3\cdot{~~76.4573}~(f_{\,1~})~\,$ & 34~~,~~~30 & 35.5700 &  0.0096 $(k_2)$ \\ 
                & $3\cdot{181.9716}~(f_{47}) - 2\cdot{255.1673}~(f_{65})~\,$ & 45~~,~202 & 35.5803 &  0.0007 $(k_2)$ \\
                & $3\cdot{158.0832}~(f_{35}) - 3\cdot{140.5215}~(f_{29})~\,$ & 173~~,~185 & 52.6851 &  0.0029 $(k_1)$ \\
                \\
                UZ & $   \bf{210.6816 ~(f_{55}) -  \,~~~~196.4174~(f_{\,51})\textsuperscript{*}}$ & 793~~,~~~36 & 14.2642 & $ \bf{0.0033~(k_3)}$ \\
                \smallskip
                & $   \bf{238.9829 ~(f_{62}) -  \,~~~~224.7191~(f_{\,59})}~\,$ & 91~~,~~~54 & 14.2638 &  $ \bf{0.0028~(k_3)}$ \\
                & $3\cdot{326.7867}~(f_{75}) -  \,~~~~966.0903~(f_{93})~\,$ & 53~~,~~~35 & 14.2698 &  0.0088 $(k_3)$ \\ 
                & $3\cdot{201.1749}~(f_{53}) - 3\cdot{196.4174}~(f_{51})\textsuperscript{*}$ & 34~~,~~~36 & 14.2726 &  0.0116 $(k_3)$ \\
                & $3\cdot{158.0828}~(f_{35}) - 3\cdot{140.5205}~(f_{29})~\,$ & 204~~,~223 & 52.6868 &  0.0045 $(k_1)$ \\
                \hline 
                \\
                \multicolumn{3}{l}{~~~~~~~~~~~~~~~~\it{\textsuperscript{*} frequency close to one of the known Kepler artifacts} }    \\    
        \end{tabular}
\end{table*}

It has previously been shown that the low-frequency signals of sdBVs  KIC\,5807616 and KIC\,10001893 cannot be combination frequencies of the stellar highest amplitude pulsating modes \citep{Charpinet2011, Silvotti2014}. We here therefore decided to calculate all pulsating mode frequency combinations independently of their amplitudes and determine whether any of them falls close to the observed low-frequency signals. The pulsation frequencies for our analysis were prewhitened from data of the two sdBVs, and for KIC\,10001893, were also taken from the recent work by \cite{Uzundag2017}. The putative exoplanetary $F_{1}$\,=\,48.204\,$\mu$Hz and $F_{2}$\,=\,33.755\,$\mu$Hz signal (frequencies and $F_{1}$ and $F_{2}$ labels according to \cite{Charpinet2011}) and $k_{1}$\,=\,52.68\,$\mu$Hz, $k_{2}$\,=\,35.58\,$\mu$Hz and $k_{3}$\,=\,14.26\,$\mu$Hz (frequencies and $k_{1}$, $k_{2 }$, and $k_{3}$ labels used to denote the exoplanetary frequencies found by \cite{Silvotti2014}) were also prewhitened from the corresponding light curves. Because we used different data sets, the resulting frequencies were not exactly the same as in the earlier publications 
(the new frequency values we found for the $F_{1}$ and $F_{2}$ signals were 48.182\,$\mu$Hz and 33.839\,$\mu$Hz, respectively). 

We took into consideration the combinations of gravity-mode frequencies $f_{i}$ and $f_{j}$ in the shape of $n\cdot{f_{i}} - m\cdot{f_{j}}$ for low integer values of $n$ and $m$. Any resulting combination 
frequencies would be of interest if they were sufficiently close to the 
observed signal in the low-frequency region. This leads to the question what 'sufficiently close' means.

Prewhitening our sets of SC data gives a frequency fit with formal errors below $\pm$0.001\,$\mu$Hz. When we allow the
\textbf{}distance of a \textbf{}combination frequency to the \textbf{}observed frequency (DCO) to be smaller than or equal to the combined and observed frequency errors alone (determined from prewhitening), then the DCO should be smaller than $\sim$0.0014\,$\mu$Hz.
However, the accuracy of a prewhitened frequency depends on the way in which the data were prepared and on the choice of a FT peak for prewhitening. The reason for this is that by adding or removing parts of the data,  we allow the pulsation FT spectrum of the star to change by adding or removing sets of the data with various pulsation activities. Because the pulsating modes can be unstable, new data sets can change the overall shapes of the modes. As a result, a prewhitened frequency might differ, depending on the data set and visibility of the pulsation mode peaks above the detection threshold.   

\begin{table*}[t]
	\caption{KIC\,10001893: Frequency amplitude modulation for the $k_{1}=52.682\,\mu$Hz frequency, prewhitened from
		LC aperture var data. Only pulsating frequencies with amplitudes greater than
		20\,ppm and matching right-side lobe amplitudes comparable to $k_1$ are shown. }
	\label{modul}      
	\centering                          
	\begin{tabular}{l r r r r r c}        
		\hline                 
		Data &      &  [$\mu$$Hz$]   &  & distance to central [$\mu$$Hz$]   & amplitudes [ppm]  & differences of side lobe \\  
		& left & central freq.~~ & right~~~~~~~~ & left~~~~~~~~~~~~~~~~~~right~~~~ & left~~central~~right & distances from central \\
		\hline                  
		\\  
		\smallskip
		LC & $\bf{k_1}$ & $ \bf{77.5306 ~(~f_{2})~}$ & $\bf{102.3475 \,~(f_{12})}$ & ~~\bf{24.8484~~~~~~~~~~24.8168~~} & \bf{22~~~~~~86~~~~~~~21}~~ & 0.032 \\ 
		& $k_1$ & $114.6705 ~(f_{18})$ & $176.6796 ~(f_{44})$ & ~~61.9883~~~~~~~~~~62.0091~~ & 22~~~~~~29~~~~~~~25~~ & 0.021\\ 
		\\
		SC & $k_1$ & $114.6710 ~(f_{18})$ & $176.6694 ~(f_{44})$ & ~~61.9888~~~~~~~~~~61.9984~~ & 22~~~~~~34~~~~~~~34~~ & 0.010\\ 
		\hline
	\end{tabular}
\end{table*}

For example, when we take frequencies determined from our SC data set and from \cite{Uzundag2017}, we find that some frequencies differ greatly between the two sets. A similar conclusion emerges when we calculate the frequency differences prewhitened from our LC data and those from \cite{Uzundag2017}.
By restricting the pulsating modes to those with amplitudes above 30 ppm, which is the $\sim$5\,$\sigma$ detection threshold of LC data (calculated within the 60-450\,$\mu$Hz FT range of the prewhitened light curve), we found that in some cases, the differences can reach nearly 0.01\,$\mu$Hz. However, the average value of all frequency differences between our data sets and those of 
\cite{Uzundag2017} is lower, approximately 0.005\,$\mu$Hz, and this value is taken as an error of a single frequency in our further considerations. For two combined frequencies, the error of a combination frequency therefore is $\pm$0.007\,$\mu$Hz and determines the maximum distance of the combined frequency to a low FT signal. 

On the other hand, the $k_{3}$ frequency of KIC\,10001893 determined from the SC aperture and LC PSF sets of light curves is the same to about $\pm$0.0001\,$\mu$Hz, while the $k_{1}$ and $k_{2}$ peak frequencies depend on the data set and can differ by as much as 0.028\,$\mu$Hz. This is more than the best FT resolution that can be achieved for LC data (0.012\,$\mu$Hz) and allows for larger DCO. The DCO clearly cannot be calculated directly from the formal prewhitened frequency errors. In the best case, the DCO should not be higher than 0.007\,$\mu$Hz. In the worst case, it can be larger than the FT resolution of the data. 
Taking into account the 0.007\,$\mu$Hz combination frequency error and the FT resolutions, we can adopt $\sim$0.019\,$\mu$Hz as the maximum DCO value for KIC\,10001893. For KIC\,5807616, the data FT resolution is 0.016\,$\mu$Hz, therefore the maximum DCO reaches $\sim$0.021\,$\mu$Hz.

Based on these DCOs values, we find that combination frequencies of intermediate-amplitude g modes can explain the $F_{1}$ and $F_{2}$ low-frequency signals of KIC\,5807616, while for KIC\,10001893, a simple beating frequency of two g modes can be solely responsible for the $k_{3}$ signal. Using the frequency list of Uzundag et al.  (2017; and their $f_{x}$ frequency indications), we can obtain the $k_3$ signal as a beating of two parent frequencies, one with a high amplitude ($f_{55}$) and the other with a low amplitude ($f_{51}$). When higher order combinations are considered, all three low-frequency signals can be explained as a result of combined pulsation frequencies that fall within $\pm{0.007}{\mu}$Hz from the $k_{1}$, $k_{2}$, or $k_{3}$ signals.
Tables \ref{combi_char} and \ref{combi_sil} summarize our efforts to determine pulsation frequency combinations that fall in the proximity of the $F_{1}$ and $F_{2}$ and  $k_{1}$, $k_{2}$, and $k_{3}$ signals for the two sdBVs. 
The last columns of the tables (called combination to ... distance) show the DCO in $\mu$Hz.

The combination frequency in the shape of $2f_{i}-f_{j}=k_{l}$ can be considered as an amplitude modulation of the $f_{i}$ peak \citep{Breger2006}. In the FT of a light curve, any frequency amplitude modulation will result in two side frequencies (lobes) of equal amplitudes that are separated symmetrically from the modulated (central) frequency. The distance of the side lobes from the central peak is equal to the modulation frequency. 
Following these assumptions, Table\,\ref{modul} shows that for KIC\,10001893, the LC and SC data FTs point to the same central peak at 114.67\,$\mu$Hz for which the amplitude could be modulated with a period of $\sim$4.5\ hours ($\sim$62.0\,$\mu$Hz). However, the amplitude of this peak is close to the 5$\sigma$ detection threshold and the peak can be disregarded. 
Allowing for larger ($\sim$0.03\,$\mu$Hz) differences of side-lobe distances from the central peak, we can find that $f_{2}$ (77.53\,$\mu$Hz, 86 ppm, LC data) frequency amplitude modulations could make two equal-amplitude side lobes. In this case, the left lobe would correspond to $k_1$ and the right lobe to $f_{12}$ (102.3475\,$\mu$Hz). The side lobe that is $\sim$24.85$\mu$Hz distant from the central peak would then correspond to a modulation period of $\sim$11.2 hours. Unfortunately, short-period modulations are difficult to measure based on the data we have. For data from \cite{Uzundag2017}, we could not find any reasonable case of amplitude modulation.

In summary, it is possible to find a beating combination of pulsation frequencies that would be responsible for the $F_{1}$, $F_{2}$ , and $k_{3}$ signals in the FT of the KIC\,5807616 and KIC\,10001893 light curves. Two other $k_{1}$ and $k_2$ signals observed in the KIC\,10001893 light curve FT can be explained only when we allow for higher order frequency combinations. Moreover, while the modulation of $f_{2}$ amplitude could in theory explain the $k_{1}$ signal, it cannot be confirmed by direct measurements of $f_{2}$ amplitude variations.  

\section{KIC\,10001893 low-frequency region}

$\it{Kepler}$ data are known for systematic artifacts that dominate certain frequencies within the time-series data. The artifacts are present with different intensities for different stars. The intensity of the artifact also depends on the number of pixels that is used to extract data. The most important artifact for the low-frequency FT region  is the reaction wheel passive thermal cycle at 3.8\,$\mu$Hz ($\it{Kepler}$ data release 12). The comb of frequencies around this artifact is usually greatly reduced in amplitude during the detrending process of the light curves. However, the artifacts and their harmonics cannot be entirely removed. The frequency amplitudes of the artifacts can also be amplified depending on the reduction methods used to receive the light curves. In our case, PSF can produce larger artifact amplitudes than aperture-reduced data.

\begin{figure}
        \centering
        \includegraphics[width=8.8cm, clip=]{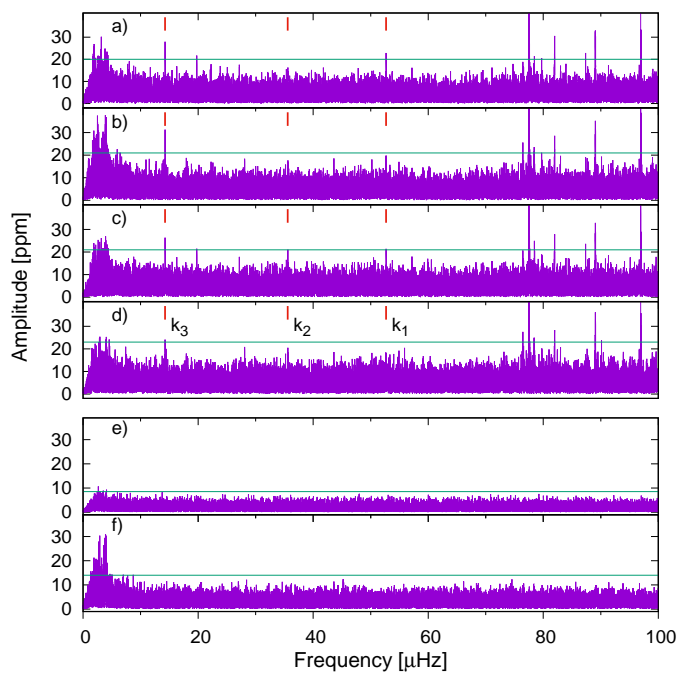}
        \caption{LC light-curve FT amplitude spectra of KIC\,10001893 (var) and
        bright neighbor (cmp) stars shown within the low-frequency region. Panel a)~ FT of var aperture reduced data ($4\sigma=20\:ppm$), b)~var PSF ($4\sigma=21\:ppm$), c)~var/cmp aperture  ($4\sigma=22\:ppm$), d)~var/cmp PSF ($4\sigma=23\:ppm$), e)~cmp aperture ($4\sigma=9.2\,ppm$), and f)~cmp PSF ($4\sigma=14.0\,ppm$). Green lines indicate the $4\sigma$ signal detection thresholds. Vertical short bars in panels a, b, c, and d denote places where $k_1 = 52.683\:\mu$Hz, $k_2 = 35.578\:\mu$Hz, and $k_3 = 14.261 \:\mu$Hz frequencies were found by \cite{Silvotti2014}.}
        \label{Fig_2}
\end{figure}

Although using a comparison star for the space data seems to be unnecessary, we demonstrate that by dividing the KIC\,10001893 light curve (var) by the light curve of the bright neighbor comparison star (cmp), the FT artifact comb amplitudes can be reduced to below 10\,$\mu$Hz. We show in Fig.\,\ref{Fig_2} the sdBV and
bright neighbor star FTs of the LC aperture and PSF light curves, which cover the low-frequency region. Green horizontal lines indicate the detection thresholds (calculated within 0\,-\,100\,$\mu$Hz), which equal 20\,ppm, 21\,ppm, 22\,ppm, 23\,ppm, 9.2\,ppm, and 14.0\,ppm for panels a, b, c, d, e, and f, respectively. Aperture and PSF data FTs for the KIC\,10001893 variable are shown in Fig.\,\ref{Fig_2} panels a and b, respectively, while var/cmp light curve FTs for aperture and PSF-reduced data are presented in panels c and d, respectively. At the bottom of Fig.\,\ref{Fig_2} we show the aperture and PSF light curve FTs of the comparison star alone.

The FTs of the aperture light curves show the lowest 3.8\,$\mu$Hz artifact comb amplitudes for the variable (panel a) and comparison stars (panel e), while FTs of PSF reduced data show amplified artifact combs (panels b and f). Similarly, a signal at the $k_3$ frequency increases its amplitude from 28\,ppm in the aperture light curve FT (panel a) up to 31\,ppm in the FT of the PSF-reduced light curve (panel b). Neither aperture nor PSF light-curve FT of the variable show signs of the $k_2$ frequency (Fig.\,\ref{Fig_2} panels a and b, respectively), while the $k_1$ amplitude drops below the detection threshold in the FT of the PSF var light curve (panel b). In addition, a new signal appears in panel a near $\sim$20\,$\mu$Hz (Table\,\ref{table_k}, labeled as $k_n$) that is not present in the FT of PSF data (panel b). 

The FT amplitude spectra of the divided var/cmp light curves show a large decrease in artifact comb amplitudes near the $\sim$3.8\,$\mu$Hz frequency for the FTs of aperture (panel c) and PSF (panel d) reduced data. Because the PSF light curves show no mutual influence (see Fig.\,\ref{Fig_1}), we focused on panels b and d. The highest $k_3$ peak of the PSF var/cmp FT decreases from $\sim$31\,ppm (panel b) to $\sim$24\,ppm, which is close to the $\sim$23\,ppm FT detection threshold shown in panel d. The decrease is about 6\,ppm, which is twice the $\sim$3\,ppm FT noise value that was introduced from dividing the variable light curve by the light curve of the nearby star. We also note an increase in the $k_2$ signal by ~2ppm in panel d, while $k_1$ decreased by ~2ppm, which is consistent with added FT noise. The peak at $\sim$28\,$\mu$Hz (visible only in PSF data) remained at the same $\sim$19\,ppm level. However, all the signals below the FT detection threshold should be disregarded. The list of peaks found above the FT detection thresholds within panels a- d is presented in Table\,\ref{table_k}.

Our study shows a strong dependence of the FT signal on the data set that is used for the analysis and on the data reduction methods that are applied. In the FT of variable aperture data, instead of $k_{2}$, a signal at 19.77\,$\mu$Hz would have to be taken into account. In the variable PSF data FT, only the $k_{3}$ signal remains strong and above the 4$\sigma$ detection threshold. Finally, after dividing the var/cmp PSF data, its light curve FT shows a weak $k_{3}$ amplitude above the detection threshold, and there is no other clear frequency signal that could be considered significant. 

\begin{table}
        \caption{Amplitudes and frequencies of FT peaks found close to the detection thresholds in panels a-d of Fig.\,2. Asterisks indicate a frequency peak that differs by more than 0.02\,$\mu$Hz from the $k_2$ frequency determined by \cite{Silvotti2014}. $k_n$ is a new frequency peak found in panels a and c of Fig.\,2. A value greater than 4 in the last column means a detection, and a value below 4.0 indicates a non-detection of a peak.}   
        \label{table_k}      
        \centering                          
        \begin{tabular}{c c c c}        
                \hline\hline                 
                ID   & Frequency   & Amplitude  & Amp/$\sigma$ \\
                & $\mu$Hz   & ppm  &   \\
                \hline
                \multicolumn{4}{l}{~~~~~~~~~~panel a ~~~var ~~~~~~~~aperture data}\\ 
                $k_3$     &14.260     &28.00     &5.56\\
                $k_1$     &52.681     &22.89     &4.55\\
                $k_n$     &19.774     &21.78     &4.32\\
                \hline
                \multicolumn{4}{l}{~~~~~~~~~~panel b ~~~var ~~~~~~~~PSF~ data}\\
                
                $k_3$     &14.260     &31.36     &5.84\\
                \hline                      
                \multicolumn{4}{l}{~~~~~~~~~~panel c  var/cmp ~~~aperture data}\\ 
                
                $k_3$     &14.260     &26.52     &4.85\\
                $k_1$     &52.682     &21.53     &3.94\\
                $k_n$     &19.774     &21.52     &3.94\\
                $*k_2$     &35.604     &21.26     &3.89\\
                \hline
                \multicolumn{4}{l}{~~~~~~~~~~panel d  var/cmp ~~~PSF~ data}\\    
                
                $k_3$     &14.260     &24.23     &4.16\\
                
        \end{tabular}
\end{table}

\section{Simulated data and KIC\,5807616}

The interpretation of  the putative exoplanetary signals observed in the sdBVs KIC\,10001893 and KIC\,5807616 is clearly challenging. Using similar sets of data, we can derive different conclusions that support alternative hypotheses about their sources. To overcome this problem, more objective methods are required to determine the source of low-frequency FT signals. One such method can be a statistical approach. For example, using simulated light curves, we could determine the FT properties of signals coming from the hypothetical exoplanets and compare them with the observed low-frequency signals. Two obvious observed parameters would be the FT frequency and amplitude, but also their time variations and the estimates of statistical errors. 

A synthetic light curve for such simulations needs to be composed of a constant flux from a star with the injected exoplanetary signal, which (in general) depends on planetary orbital parameters that include the inclination $i$ toward the observer. The light curves that are obtained in this way are usually highly non-sinusoidal \citep{KaneGel2011}, and their FTs show harmonic frequencies that correspond to the frequency of the exoplanetary orbital period. Unfortunately, harmonics have much lower amplitudes than their parent frequencies, and any information about harmonics (or light-curve shapes) is lost in the FT noise. However, the frequency and amplitude of the brightness variation due to orbital motion can be retrieved with an accuracy that is proportional to the $\sin$ $i$.

When orbits around the host stars become very tight, we expect them to be circularized, and the exoplanetary fluxes will add only sinusoidal components of light variability to the stellar luminosities. As follows for Eq. D.11 in \cite{Charpinet2011}, in this case, the luminosity variations due to exoplanetary reflected and radiated flux take the form
 \begin{equation}
 \begin{split}
 \frac{L_p}{L_{\star}} =\left\{\frac{A_{g}} 8\frac{R_p^2}{a^2}+2\pi R_p^2\frac{F_R(T_p)-F_R(\beta T_p)} {L_{\star}}\right\}\sin i\sin
 \frac {2 \pi t } {P_p}
 \end{split}
 ,\end{equation}
where $L_\star$ is the luminosity of the host star, $A_{g}$ is the Bond albedo, $F_R(T_p)$ is the total radiated exoplanetary flux convolved with the $\it{Kepler}$ filter. $T_p$ and $\beta T_p$ denote equilibrium temperatures of the hot and cold sides of an exoplanet, respectively. The $\beta$ parameter describes the temperature contrast between the hot and cold sides of the exoplanet (it can take values between 0 and 1). The other exoplanetary parameters are its radius $R_{p}$, the luminosity $L_{p}$, the orbital period $P_p$ , and the distance to the host star $a$. 
As a result, the relative flux variations are strictly periodic and of a constant amplitude proportional to the $\sin$ $i$. When we neglect changes in the exoplanetary flux introduced by other exoplanets and moons in the system (or the weather in an exoplanet atmosphere, \cite{Bear2014}), the simulated light curve takes the form $L(t) = const.+ A \sin(2 \pi t/P_p)$, where $A$ is the amplitude of the exoplanetary flux variation and $t$ is time.
  
\begin{figure}
\centering
\includegraphics[width=8.8cm]{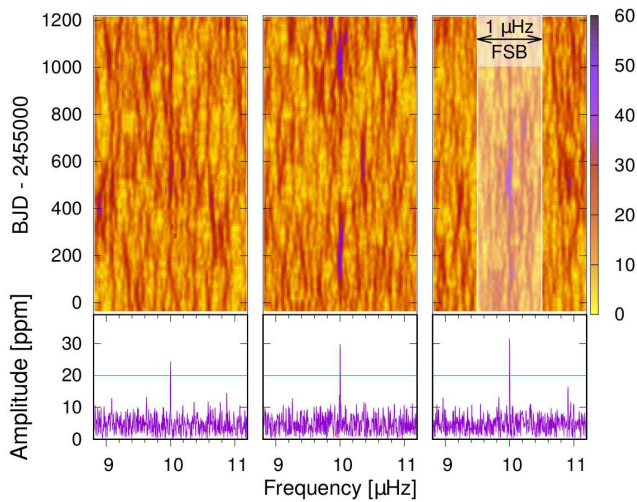}
\caption{Three examples of the 200-day running FT (top panel) and FT of the complete simulated light curves (bottom panel). The amplitudes and frequencies of the simulated signals were set at 10\,$\mu$Hz and 30\,ppm. Green lines in the bottom panel represent 4$\sigma$ detection thresholds. A frequency search box of 1.0\,$\mu$Hz width, used to find FT signals (see text) is shown in the top right panel as the shaded white and transparent rectangle.}
\label{Fig_3}
\end{figure}       

The continuous L(t) function was sampled and binned (with multiple 6.02\,s integration times followed by 0.52\,s readouts) to mimic the original 29.42-minute LC $\it{Kepler}$ data. The binned points were distributed in time in the same way as the original data (including gaps), and scattered to reflect the photon noise observed in $\it{Kepler}$ light curves. Then the FT and the running FT were calculated from a simulated light curve. In Fig.\,\ref{Fig_3} we present the examples of a 200-day box with the 10-day shift running FT in the top panel and the FT of an entire simulated LC light curve in the bottom panel, mimicking Q\,1-Q\,17 LC quarters. The signal frequency and amplitude were set to 10\,$\mu$Hz and 30\,ppm each time. The photon noise in the light curves was generated at a level equivalent to 5\,ppm noise in the FT of these light curves (calculated between 0 - 60 $\mu$Hz). This is equivalent to a $\simeq$\,20\,ppm (4$\sigma$) detection threshold of the whole (1440 days) light-curve FT and to a 50-70\,ppm detection threshold (depending on a quarter of the data) of the 200-day running FT. The signal amplitude clearly varies in the running FT panels of Fig.\,\ref{Fig_3} depending on the light-curve draw. We do not observe frequency splitting due to the noise in the top or bottom panels. 

\begin{figure}
        \centering
        \includegraphics[width=8.8cm]{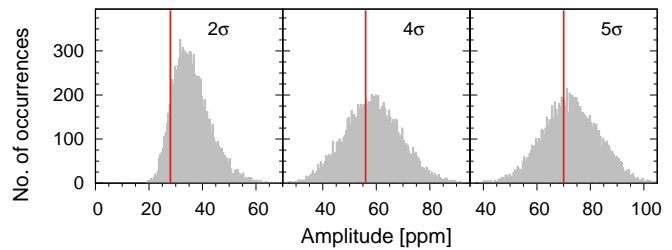}
        \caption{Signal amplitude distributions in 200-day FTs of simulated light curves.}
        \label{Fig_4}
\end{figure}

The properties of low-frequency signals were derived from their amplitude and frequency distributions out of 10\,000 light-curve simulations. Each time a light curve FT was calculated, the maximum peak was searched for within a frequency search box (FSB) $\pm0.5\mu$Hz around the simulated frequency (Fig.\ref{Fig_3}). The peak amplitudes (and frequencies) were then used to prepare the histograms of the FT amplitude (and frequency) distributions. The histograms for the 200-day light-curve amplitude are shown in Fig.\,\ref{Fig_4}. The light curves were simulated for a constant frequency and amplitudes set at levels corresponding to two, four, and five times the average FT noise marked as $\sigma$ (which corresponds to $\simeq$\,14\,ppm). The histogram for the weakest signal (left panel) is highly asymmetric. This is due to the selection effect that is caused by a non-infinitesimal width of the FSB. For the wider FSBs, there is a higher probability that a spurious FT noise signal will be taken into account. 
The asymmetry disappears when the simulated amplitudes reach an FT detection threshold higher than 4$\sigma$. This occurs because the method of searching for the highest amplitude FT peaks only reacts to signals that come from the simulated light curves, not from the spurious FT noise. 

    \begin{figure}
\centering

\includegraphics[width=8.8cm, clip=]{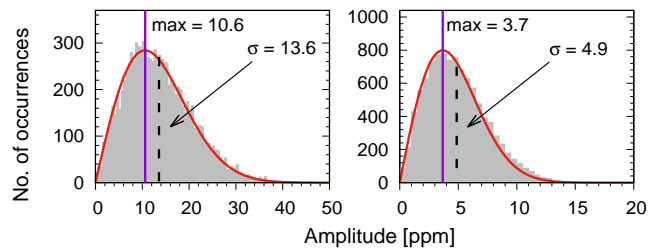}
\caption{Distributions of the noise amplitude in FTs of 200-day (left) and 1440-day (right) light curves. The simulated light-curve noise was set at the levels corresponding to $\sim$13.6 ppm of the average FT noise ($\sigma$) for the 200-day and $\sim$4.9 ppm for the 1440-day data sets. The FSB width was set to 0\,$\mu$Hz. The vertical blue lines at 10.6 and 3.7 ppm denote histogram maxima, while the average FT noise is indicated by the dashed lines.}
\label{Fig_6}
\end{figure}
\begin{figure}
\centering
\includegraphics[width=8.8cm]{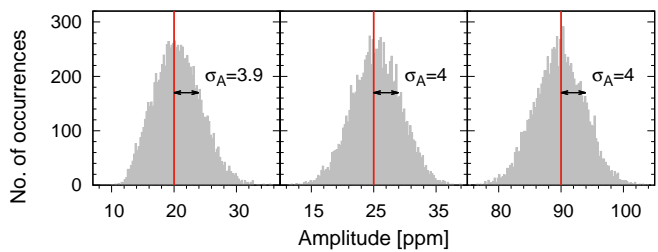}
\caption{Histograms of the FT signal amplitude distributions of the 1440-day light curves and the simulated (red vertical lines) amplitudes at 20 (left), 25 (center), and 90\,ppm (right). The amplitude scatter marked by $\sigma_{A}$ in each panel equals the maximum FT noise distribution in the right panel of Fig.\,\ref{Fig_6}.
}
\label{Fig_7}
\end{figure}

Figures\,\ref{Fig_6} and\,\ref{Fig_7} present the pure FT  noise and the amplitude distributions of the FT signal, respectively. The noise histograms were calculated for 200-day (left) and 1440-day (right) data sets with an FSB width of 0\,$\mu$Hz (i.e., measured at a single frequency). The amplitude distributions were calculated for the 1440-day light curve with an FSB width of 1\,$\mu$Hz and a signal amplitude greater than the 4, 5, and $\sim$18\,$\sigma$ FT detection thresholds. All histograms in Fig.\,\ref{Fig_7} have the same Gaussian-like shape and the same scatter $\sigma_{A}$. This scatter corresponds to the maximum FT noise amplitude distribution (MNAD) that is shown in the right panel of Fig.\,\ref{Fig_6}. The MNAD for the 1440-day data is near 4\,ppm, while the noise average is $\sim$5\,ppm. For the 200-day light-curves, MNAD is near 11\,ppm, while the average noise level is at 14\,ppm. In all cases reported here, the simulated amplitudes are recovered from the light-curve FTs with about the same MNAD accuracy that does not depend on the values of the simulated amplitude (shown as the red vertical lines in Fig.\,\ref{Fig_7}).

\begin{figure}
        \centering
        
        \includegraphics[width=8.8cm, clip=]{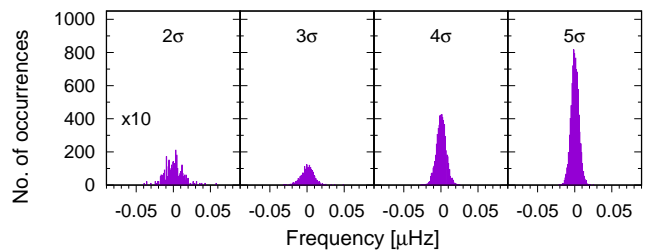}
        \caption{Signal frequency distribution in FTs of the 200-day light curves, including the constant-frequency sinusoidal component. The amplitudes of the sinusoidal components in the light curves were generated at two, three, four, and five times the $\sigma$ levels (i.e., the average FT noise value of $\sim$13.6 ppm). Only signals with amplitudes above the 4\,$\sigma$ FT detection threshold were taken into account. For high simulated amplitudes, the corresponding FT signal frequencies are confined in a narrow $\pm$0.02\,$\mu$Hz range (right panel). The range increases up to $\pm$0.03\,$\mu$Hz for the low-amplitude signals. The left panel was enlarged ten times to account for the small number of signal detections above 4\,$\sigma$.}
        \label{Fig_8}
\end{figure}

While the approximate MNAD scatter in the FT amplitudes around the values of the simulated amplitude were expected, the distributions of the FT signal frequency were not. In Fig.\,\ref{Fig_8} we show the histograms of the FT frequency distributions for the simulated 200-day light curves with amplitudes set at the 2, 3, 4, and 5$\sigma$ (here again, $\sigma$ is the FT average noise), but the frequencies were set to the same constant value. The histograms were centered on 0\,$\mu$Hz by subtracting the simulated frequency values (note the negative numbers on the x-axis). The FSB was the same as we used to calculate the amplitude histogram ($\pm$0.5\,$\mu$Hz), but only the maximum-amplitude FT peaks above 4$\sigma$ FT detection threshold were considered to avoid counting spurious FT signals. 
We note that the resulting FT frequency distributions are symmetrically distributed around the simulated light-curve frequencies. For the higher simulated amplitudes, the FT frequency distributions are narrower, that is, below $\pm$0.02\,$\mu$Hz (amplitudes at 5$\sigma$). However, the frequency scatter is still contained within $\pm$0.03\,$\mu$Hz at amplitudes near 2$\sigma$ (note that the scale in the y-axis of the left panel is ten times larger). 
This observation shows that the resulting FT signal frequency recovers the frequency we used for the light-curve simulations with high accuracy. Larger variations in the running FT signal frequency could indicate a different source of light than from an exoplanet on a stable orbit. For example, pulsation-frequency phase changes are commonly observed in sdBVs and pulsating white dwarfs (see sdBV KIC\,10670103, \citet{Krzesinski2014}; or GD358 DBV, \citet{Kepler2003}.

\begin{figure}
        \centering
        
        \includegraphics[width=8.8cm, clip=]{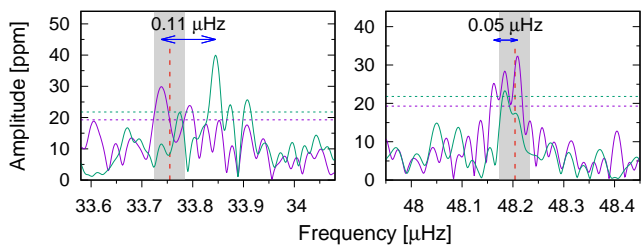}
        \caption{Fourier transforms of two halves ($\sim$700days) of KIC\,5807616 data around the $F_{2}$ (33.839$\mu$Hz) and $F_{1}$ (48.182$\mu$Hz) frequencies from \cite{Charpinet2011}. The green line shows the FT of the Q\,5 - Q\,10 light curves, and the pink line plots the FT of the Q\,11 - Q\,17 light curves.  The detection thresholds are plotted with dashed horizontal lines in colors corresponding to FT colors. The gray vertical stripes indicate the $\pm$0.03\,$\mu$Hz frequency scatter region, and the red vertical lines denote the positions of the $F_2$ and $F_1$ signals found by \cite{Charpinet2011} (see the text for details).}
        \label{Fig_9}
\end{figure}

This new FT-frequency stability test is not conclusive for stable signals such as from KIC\,10001893 $k_3$. However, it has been successfully performed on the KIC\,5807616 $F_1$ and $F_2$ frequencies, which appeared to vary in a wider frequency range \citep[see Fig.\,7,][]{Krzesinski2015}. To test our method, we divided the data of KIC\,5807616 into two light curves of $\text{about }$700 days and calculated their FTs. The FT amplitudes of the $F_1$ and $F_2$ signals prewhitened from these data sets were above the 4$\sigma$ FT detection threshold and the FT resolution given by 1.5/T, where T\, denotes the data length, was $\sim$0.025\,$\mu$Hz for both sets. The signals near $F_1$ and $F_2$ in each FT set were found and the results are presented in Fig.\,\ref{Fig_9}. The allowed 0.03\,$\mu$Hz frequency scatter region is indicated as the vertical gray stripe, while the dashed red lines indicate the original positions of the $F_1$ and $F_2$ signals from \cite{Charpinet2011}. The FT of the first half of the data (Q\,5 - Q\,10) is shown in pink, and the second (Q\,11 - Q\,17) is plotted in green.
 
The large 0.11\,$\mu$Hz separation of the $F_2$ signals in our two data sets is clearly visible (see the left panel of Fig.\,\ref{Fig_9}). The separation of the $F_{1}$ signals is less evident, but the range of signal variation in the first part of the data (pink) is larger than the frequency scatter allowed for the putative exoplanetary signals.
We conclude that the $F_2$ frequency is not the result of a stable sinusoidal light curve of the exoplanetary origin. The evidence for the non-exoplanetary source of the $F_1$ signal is weaker, but it did not pass our FT frequency stability test either. Because both signals can be beatings of the sdBV g modes with moderate amplitudes (Table\,\ref{combi_char}), we are inclined to classify them as the combination frequencies of the stellar pulsations.    

\section{Conclusions}

The previously published low-frequency signals found in the pulsating sdBV stars KIC\,5807616 and KIC\,10001893 \citep{Charpinet2011,Silvotti2014} were interpreted as of exoplanetary origin. Both stars were presented as candidates for exoplanetary systems. 

Based on our new PSF data that we extracted from the bright neighbor of KIC\,10001893, we have shown that the $k_{1}$, $k_{2}$, and $k_{3}$ signal amplitudes that are visible in low-frequency data strongly depend on the data reduction techniques. The $k_{1}$ and $k_{2}$ signals were classified as non-detectable or spurious. 
We also suggest that for the low-frequency region, the FT detection threshold should be increased to at least 5$\sigma$ to avoid including weak random signals. While we would not consider $k_{1}$ and $k_{2}$ as real signals, $k_{3}$ has passed our higher FT detection threshold test. 
However, by using a comparison star to the variable star data, we highly reduced the FT light-curve artifacts along with the amplitude of $k_{3}$. This implies that the $k_{3}$ signal is correlated with the artifacts. On the other hand, we have also shown that the $k_{3}$ frequency alone of the three frequencies that can be explained as a beating frequency of the stellar pulsating modes (Table\,\ref{combi_sil}). This means that we cannot be certain if this signal should be interpreted as an artifact or a real signal caused by stellar pulsations.

In our simulations we discovered that the simulated exoplanetary sinusoidal light curve (of a constant amplitude and frequency) gives an FT signal that is contained within a narrow $\pm$0.03\,$\mu$Hz bandpass around the simulated frequency. This feature allows us (in some cases) to distinguish between the exoplanetary and non-exoplanetary signals in the low-frequency FT regions. 

While the FT frequency stability test cannot be used to distinguish between the natures of the $k_3$ signal because of its stability in frequency, we performed this test on the $F_1$ and $F_2$ signals that were observed in KIC\,5807616 sdBV. Both signals have larger frequency variations than expected from the exoplanetary origin. After a careful study, we classified the $F_1$ and $F_2$ frequencies as resulting from a beating of intermediate-amplitude pulsating g modes.

Our simulations also showed that a good estimate for low-frequency signal amplitude errors is MNAD (Fig.\,\ref{Fig_7}). This can be used for the formal estimates of the exoplanetary size erros, if derived from the FTs of light curves.

Finally, we comment on the orbital resonance argument raised by \cite{Charpinet2011} and \cite{Silvotti2014}, who concluded that periods corresponding to the KIC\,5807616 $F_1$ and $F_2$ or the KIC\,10001893 $k_1$, $k_2$ , and $k_3$ frequencies are in resonance, which supports the exoplanetary nature of the signals. However, our analysis showed that these signals cannot be of exoplanetary origin. In addition, we refer to \cite{Staff2016}, who showed that for a planet with about $ \text{ten}$ Jupiter masses it takes about three years to spiral down onto a parent RG star if the planet is engulfed in the stellar envelope.
Although the simulations in \cite{Staff2016} are likely not final and were performed for 
a zero-age main-sequence mass of 3.5\,$M_{\sun}$, they present difficulties that they encountered when they tried to explain the survival of planets within RG star envelopes.
On the other hand, if the signals observed in KIC\,5807616 and KIC\,10001893 are combination frequencies of stellar pulsating modes, perhaps they can be interpreted as g modes that combine into a specific ratio of beating frequencies.

\cite{Silvotti2014} also argued that the $k_1$ signal might be responsible for the tidally induced frequency at 210.72\,$\mu$Hz. This high-amplitude $l\,=\,1$ pulsation g mode \cite[$f_{55}$ as identified by][]{Uzundag2017} would correspond to the third harmonic of $k_{1}$. We found, however, that a 158.08\,$\mu$Hz $l\,=\,1$ g mode \cite[$f_{35}$ in][ a few times lower in amplitude than $f_{55}$]{Uzundag2017}, fits $k_1$ better as a second harmonic.
We do not observe any unusual amplitude amplification in $f_{55}$ or $f_{35}$ , as we would expect from tidally powered modes. They both seem to fit the commonly observed sequence of $l\,=\,1$ modes in sdBV stars. Unfortunately, none of the sdBV pulsation models can predict values of pulsation amplitudes. Therefore tides cannot be proved or refuted as a source of mode excitation.

\begin{acknowledgements}
      This work was supported by the National Science Centre, Poland. The project registration number is 2017/25/B/ST9/00879.
\end{acknowledgements}


\bibliographystyle{aa}
\bibliography{35003_final.bbl}

\end{document}